\documentclass[a4paper, aps, prc, showpacs, superscriptaddress, preprintnumbers, nofootinbib, twocolumn]{revtex4-1}
\usepackage[pdftex]{graphicx, color}
\usepackage{amsmath, amssymb, amscd, latexsym, bm}
\usepackage[colorlinks=true,pdfstartview=FitV,linkcolor=blue,citecolor=magenta,urlcolor=blue,bookmarks=true,bookmarksnumbered=true]{hyperref}

\begin{document}

\preprint{YITP-20-07}

\title{Signatures of the vortical quark-gluon plasma in hadron yields}

\author{Hidetoshi Taya}
\affiliation{Department of Physics and Center for Field Theory and Particle Physics, Fudan University, Shanghai, 200433, China}
\affiliation{Research and Education Center for Natural Sciences, Keio University 4-1-1 Hiyoshi, Kohoku-ku, Yokohama, Kanagawa 223-8521, Japan}
\author{Aaron Park}
\affiliation{Department of Physics and Institute of Physics and Applied Physics, Yonsei University, Seoul 03722, Korea}
\author{Sungtae Cho}
\affiliation{Division of Science Education, Kangwon National
University, Chuncheon, 24341, Korea}
\author{Philipp Gubler}
\affiliation{Advanced Science Research Center, Japan Atomic Energy Agency, Tokai, Ibaraki 319-1195, Japan}
\author{Koichi Hattori}
\affiliation{Yukawa Institute for Theoretical Physics, Kyoto University, Kyoto, 606-8317, Japan}
\author{Juhee Hong}
\affiliation{Department of Physics and Institute of Physics and
Applied Physics, Yonsei University, Seoul 03722, Korea}
\author{\\Xu-Guang Huang}
\affiliation{Department of Physics and Center for Field Theory and Particle Physics, Fudan University, Shanghai, 200433, China}
\affiliation{Key Laboratory of Nuclear Physics and Ion-beam Application (MOE), Fudan University, Shanghai 200433, China}
\author{Su Houng Lee}
\affiliation{Department of Physics and Institute of Physics and Applied Physics, Yonsei University, Seoul 03722, Korea}
\author{Akihiko Monnai}
\affiliation{KEK Theory Center, Institute of Particle and Nuclear Studies, High Energy Accelerator Research Organization (KEK), 
Tsukuba, Ibaraki 305-0801, Japan}
\affiliation{Department of Mathematical and Physical Sciences, Japan Women's University, Tokyo 112-8681, Japan}
\author{Akira Ohnishi}
\affiliation{Yukawa Institute for Theoretical Physics, Kyoto University, Kyoto, 606-8317, Japan}
\author{Makoto Oka}
\affiliation{Advanced Science Research Center, Japan Atomic Energy Agency, Tokai, Ibaraki 319-1195, Japan}
\author{Di-Lun Yang}
\affiliation{Faculty of Science and Technology,  Keio University, Yokohama 223-8522, Japan}

\collaboration{ExHIC-P Collaboration}\noaffiliation

\begin{abstract}
We investigate the hadron production from the vortical quark-gluon plasma created in heavy-ion collisions.  Based on the quark-coalescence and statistical hadronization models, we show that total hadron yields summed over the spin components are enhanced by the local vorticity with quadratic dependence.  The enhancement factor amounts to be a few percent and may be detectable within current experimental sensitivities.  We also show that the effect is stronger for hadrons with larger spin, and thus propose a new signature of the local vorticity, which may be detected by the yield ratio of distinct hadron species having different spins such as $\phi$ and $\eta'$.  The vorticity dependence of hadron yields seems robust, with consistent predictions in both of the hadron production mechanisms for reasonable values of the vorticity strength estimated for heavy-ion collisions.  
\end{abstract}

\maketitle

{\it Introduction}.  Relativistic heavy-ion collisions (HIC) at the BNL Relativistic Heavy Ion Collider (RHIC) observed the global spin polarization of $\Lambda$ hyperons produced from the quark-gluon plasma (QGP) \cite{sta07, sta17, sta18}.  The observed spin polarization indicates that a strong vorticity field is globally generated in noncentral collisions.  The strength of the global vorticity is estimated as $\omega \!\approx\! 9 \!\times \!10^{21}\;{\rm s}^{-1} \!= \!{\mathcal O}(1\;{\rm MeV})$ \cite{sta17}, which surpasses all the other known vorticity strengths in the current Universe by orders of magnitude.  Complementary measurements with light mesons have been performed more recently at the CERN Large Hadron Collider (LHC) as well as RHIC \cite{Abelev:2008ag, Acharya:2019vpe}.  It, therefore, provides a unique and novel opportunity for studying systems with strong vorticity.  Together with related theoretical developments such as spin-transport quantum kinetic theories \cite{yan18, Mueller19, wei19, gao19, hid19, Wang:2019moi, yee19, Yang20, Liu20, Bhadury20} and relativistic spin hydrodynamics \cite{flo18, woj18, mon19, hat19}, this has created synergies among various research fields in physics including nuclear physics, astrophysics, and condensed-matter physics, in particular, spintronics.  Thus, there is growing interest in the vortical QGP and the consequent spin-dependent observables in HIC. 

Since the direction and distribution of the vorticity in the QGP depend on spacetime and its average appears as the global vorticity, the local strength of the vorticity can be greater than the global one.  Indeed, various phenomenological studies of the QGP such as hadron transport models \cite{hui17, xia18, jia17, wei18}, hydrodynamics \cite{kar17, bec18, flo19, wan19, xie19, cse19}, and holography \cite{ban18} confirmed that the local vorticity is stronger than the global one by an order of magnitude, $\omega \!\approx\! 0.3 \!\times\! T \!\approx\! 50\;{\rm MeV}$ with $T$ being the temperature.  The strong local vorticity can give rise to novel observables other than the global $\Lambda$ polarization.  An example is the azimuthal-angle dependence of the $\Lambda$ polarization along the beam direction, which was observed in Ref.~\cite{sta19}.  However, a comparison between the experimental and theoretical results has provoked a ``sign problem'' with opposite signs of measured data and theory \cite{bec18, xia18, sun19, xia19, cao19}.  Currently, more studies are demanded to deepen our understanding of the local vorticity field in HIC.

In this Rapid Communication, we propose that hadron yields may serve as a new observable for the local vorticity in HIC.  We demonstrate that (i) hadron yields summed over the spin components (dubbed as {\it spin-summed hadron yields}) depend on the local vorticity, and that (ii) the dependence is stronger for hadrons with larger spin.  Thus, a systematic measurement of spin-summed hadron yields (of any hadron species) and their spin-size dependence may bring about a signature of the local vorticity.  This is complementary to the current polarization measurements, which are based on the analysis of final-state momentum distributions of hadrons decaying from particular species of hadrons such as $\Lambda, K^{\ast0}$, and $\phi$ \cite{sta07, sta17, sta18, Abelev:2008ag, Acharya:2019vpe}.  Our proposal also implies that hadron yield predictions are improved by taking into account of the finite local vorticity in traditional hadronization models.

Our basic idea is the following: Within the quark model, the hadron spin is composed of the addition of the constituent quark spins.  Therefore, particular hadron spin states, which are composed of aligned quark spins, are favorably produced against hadrons in the other spin states if quarks in the QGP are spin polarized \cite{XNWang, lia04, lia05}.  {\it This spin-selection effect in hadron production survives even in spin-summed hadron yields}.  The modification to spin-summed hadron yields should depend only on even powers of the  vorticity $\omega$ (or, generally, its strength).  Hence, spin-summed hadron yields do not suffer from the cancellation among local vorticity fields with fluctuating directions when integrated over the freeze-out surface.

To demonstrate our idea, we first construct a phenomenological hadron production model by extending the quark-coalescence model (see, e.g., Ref.~\cite{fri08}).  We assume that constituent quarks (whose mass is, e.g., $\!\approx\! 300\;{\rm MeV}$ for up and down quarks) have a local thermal distribution at the coalescence.  The energy shift by a spin-vorticity coupling \cite{bec12, bec13, vil80, lan80, mat11, heh90} induces quark-spin polarization.  We estimate hadron yields to be proportional to products of the spin-polarized quark distributions.  Second, to estimate possible model dependences, we compare the quark-coalescence model with the established statistical hadronization model (see, e.g., Ref.~\cite{bra04}).  This model does not contain quark degrees of freedom, and vorticity couples to hadron spin rather than quark spin.

We show analytically and numerically that, in spite of the aforementioned difference in the spin-vorticity coupling of the two models, spin-summed yields of $S$-wave hadrons are (almost) {\it model independently} enhanced by vorticity by a factor $1\!+\!s(s\!+\!1)(\omega/T)^2/6$, where $s$ is the hadron spin, for reasonable values of vorticity strength estimated for HIC.  The model independence implies the robustness of the vorticity effects on hadron yields; they are not strongly affected by the hadron production mechanisms.  The enhancement suggests that the strong local vorticity in HIC $\omega/T ={\mathcal O}(0.1)$ modifies hadron yields by a few percent $(\omega/T)^2 \!=\!{\mathcal O}(1\;\%)$.  This may be observable within the current experimental sensitivities.

{\it Quark-coalescence model}.  We assume that quarks have a local thermal distribution at the coalescence.  Due to a spin-vorticity coupling, the quark energy is modified by $\delta E \!=\! - \omega s_z$, where $s_z\!=\!+1/2$ ($-1/2$) for spin up $\uparrow$ (down $\downarrow$) is the spin component along the vorticity.  The spin-dependent (anti)quark distribution $n_s$ reads
\begin{align}
    n_{s}(\omega)
       	&:= \int \frac{{\rm d}^3{\bm x}{\rm d}^3{\bm p}}{(2\pi)^3} 
       	\frac{1}{ {\rm e}^{  E_{\rm q} /T } + 1  } 
		= n(\omega) \frac{1 \pm P(\omega)}{2},  \label{eq-1}
\end{align}
where $E_{\rm q}\!:=\!\sqrt{m_{\rm q}^2\!+\!{\bm p}^2} \!-\!  {\bm \mu} \!\cdot\! {\bm Q}_{\rm q} \!-\! s_z \omega$.  $m_{\rm q}, {\bm p}$, and ${\bm Q}_{\rm q}$ are the mass, momentum, and charges (i.e., electric charge, baryon number, etc.) of a (anti)quark, respectively.  We define the total quark number and the quark polarization as $n \!:=\! n_{\uparrow} \!+\! n_{\downarrow}$ and $P \!:=\! [n_{\uparrow} \!-\! n_{\downarrow}]/n$, respectively.  We assume that the quark coalescence takes place at temperature $T$, and $\omega$ and ${\bm \mu}$ are the corresponding values for vorticity and chemical potentials, respectively.  Generally, one can introduce the space-time dependence of $T$, ${\bm \mu}$, and $\omega$, but we assume for simplicity that they are given in the local rest frame and take a constant value.

Subsequently, the polarized quarks coalesce to form a hadron. We remark that (i) only a particular combination of quark spins is allowed to form a hadron, and thus the quark polarization $P$ should affect the probability of coalescence to form a hadron, which is taken into account by a spin combinatorial factor denoted by $C(P)$ below; and that (ii) hadron yields should be proportional to the total number of each constituent quark $n$, i.e., $N^{\rm coal} \!\propto\! \prod_{q={\rm quarks}} n_q$, which is a general feature of the quark-coalescence model.  Hence, hadron yields $N^{\rm coal}$ depend on vorticity via the $\omega$ dependence of $P$ and $n$ as
\begin{align}
    \frac{N^{\rm coal}(\omega)}{N^{\rm coal}(\omega=0)}
    = \prod_{q={\rm quarks}} \frac{n_{q}(\omega)}{n_{q}(0)} C(P).   \label{eq-2} 
\end{align}
The value of the spin combinatorial factor $C(P)$ is
\begin{subequations}
\begin{align}
	C_{{\rm meson}} 
		&=	\left\{ 
				\begin{array}{ll} 
					1-P^2 & (s=0) \\ 
					1+P^2/3  & (s=1)
				\end{array} 
			\right. , \label{eq-3a}\\
	C_{{\rm baryon}} 
		&=	\left\{ 
				\begin{array}{ll} 
					1-P^2 & (s=1/2) \\ 
					1+P^2 & (s=3/2)
				\end{array} 
			\right. , \label{eq-3b}
\end{align}
\end{subequations}
for $S$-wave hadrons.  $P$ here is assumed to be independent of quark flavors, which is a higher order effect.  For example, to form a proton with spin up and down, quarks must have spin combination $(\uparrow, \!\uparrow, \!\downarrow)$ and $(\uparrow, \!\downarrow,\! \downarrow)$, respectively.  As the probability to find a quark with spin up and down is $(1\!+\!P)/2$ and $(1\!-\!P)/2$, respectively, the probability $C_{\uparrow/\downarrow}$ to form a proton with spin up/down reads $C_\uparrow \!=\! (1\!+\!P)^2(1\!-\!P)/2$ and $C_\downarrow \!=\! (1\!-\!P)^2(1\!+\!P)/2$.  Thus, $C \!=\! C_\uparrow \!+\! C_\downarrow \!=\! 1\!-\!P^2$.  Naturally, no protons are produced if quarks are completely polarized $|P|\!=\! 1$.  Note that we normalized $C$ against the polarized quark distribution, i.e., $C(P\!\!=\!\!0)\!\!=\!\!1$.  Hence, the additional factors of $n_q(\omega)$ in Eq.~(\ref{eq-2}) are needed so that the coalescence production is proportional to the quark numbers at finite vorticity.

$N^{\rm coal}$ should be invariant under interchange between the quark spin labels (or a flip of the vorticity direction), while $P$ ($n$) is odd (even) under this transformation by definition.  Consequently, $N^{\rm coal}$ should depend only on the absolute value of $P$.  This general observation is confirmed with the above $C$ factors, which indeed have only even dependences on $P$.  This implies that the leading vorticity correction is quadratic, i.e., $ N^{\rm coal} \!=\! 1\!+\!{\mathcal O}(\omega^2)$ as it is evident $ n(\omega)\!-\! n(0) \!=\!  {\mathcal O}( \omega^2) $ and $P \!=\! {\mathcal O}(\omega)$ from their definitions.  In more general, $ N^{\rm coal} $ could be nonanalytic in $|\omega|$ if the nonperturbative and/or off-equilibrium hadronization process generates nonanalytic dependences on the initial conditions.

{\it Statistical hadronization model}.  We assume that vorticity couples to hadrons as $\delta E \!=\! -s_z \omega$ ($s_z \!=\! -s,\! -s+1,\! \ldots,\! +s$ with $s$ being hadron spin).  Hadron yields at the chemical freeze-out are determined by the thermal hadron distribution modified by this energy shift.  For $S$-wave hadrons, we have (cf. Ref.~\cite{bec17})
\begin{align}
	N^{\rm stat}
	&:= 
	\sum_{s_z = -s}^{ +s} 
	\int \frac{{\rm d}^3{\bm x}{\rm d}^3{\bm p}}{(2\pi)^3} 
	\frac{1}{ {\rm e}^{ E_{\rm h}/T^{\rm ch}} \mp 1 }, \label{eq-5} 
\end{align}
where $E_{\rm h} \!=\! \sqrt{m^2_{\rm h}\!+\!{\bm p}^2} \!-\! {\bm \mu}^{\rm ch} \!\cdot\! {\bm Q}_{\rm h} \!-\! \omega^{\rm ch} s_z $ and the minus (plus) sign for mesons (baryons).  $m_{\rm h}$, ${\bm p}$, and ${\bm Q}_{\rm h}$ are the hadron mass, momentum, and charges, respectively.  We assume the chemical freeze-out to take place at temperature $T^{\rm ch}$, at which the chemical potentials and vorticity are given by ${\bm \mu}^{\rm ch}$ and $\omega^{\rm ch}$, respectively.  Note that, in case of the statistical hadronization model, it is already clear in Eq.~(\ref{eq-5}) that the hadron yield $N^{\rm stat}$ is even in $\omega$.

The coalescence and chemical freeze-out temperatures, $T$ and $T^{\rm ch}$, can in general be different from each other.  Nevertheless, the difference is expected to be small such that it does not change our results significantly.  For example, theoretical analyses show that $|T\!-\!T^{\rm ch}|/T \!\approx\! 2\;\%$ at RHIC \cite{exh17}, and that the vorticity decays only slowly at late times \cite{wei18, jia17, hui17, xia18, kar17, cse19}.  Thus, we may assume $X^{\rm ch} \!=\! X \;(X\!=\!T,{\bm \mu}, \omega)$, and drop the superscript below.

{\it Boltzmann approximation}.  To understand the $\omega$ dependence of hadron yields, we analytically evaluate the two models with the Boltzmann distribution.  This is a good approximation since hadron and quark masses except for pion are significantly heavier than the typical coalescence/chemical freeze-out temperature at HIC $\!\approx\!160\;{\rm MeV}$ and thus $\exp(E_{\rm q,h}/T) \!\gg\! 1$ holds.  After expanding the Boltzmann distribution in terms of $\omega$, we find that the two models give exactly the same result
\begin{align}
	\frac{N^{\rm stat/coal}(\omega)}{N^{\rm stat/coal}(\omega=0)} \sim  1 + \frac{s(1+s)}{6} \left( \frac{\omega}{T} \right)^2 \label{eq-7}
\end{align}
for mesons (baryons) $s\!=\!0,1$ ($s\!=\!1/2,3/2$).  Thus, hadron yields increase quadratically with $\omega$, and hadrons with larger spin are more strongly enhanced.  The origin of the agreement is traced back to the fact that $s_z$ of a hadron is the sum of those for individual quarks, implying that the change of hadron yields by vorticity is less affected by the hadron production mechanisms.  This argument is reasonable as long as vorticity is weak $\omega \!\lesssim\! m_\pi$, for which the inner quark structure of hadrons is less important.

Equation~(\ref{eq-7}) is an even function of $\omega$ as we foresaw from the invariance with respect to the interchange of the quark spin labels.  This means that spin-summed hadron yields are independent of the direction of vorticity.  Hence, spin-summed hadron yields are free from the cancellation among local vorticity fields with fluctuating directions when integrating over the freeze-out surface.

\begin{figure}
\begin{center}
\includegraphics[width=0.9\linewidth]{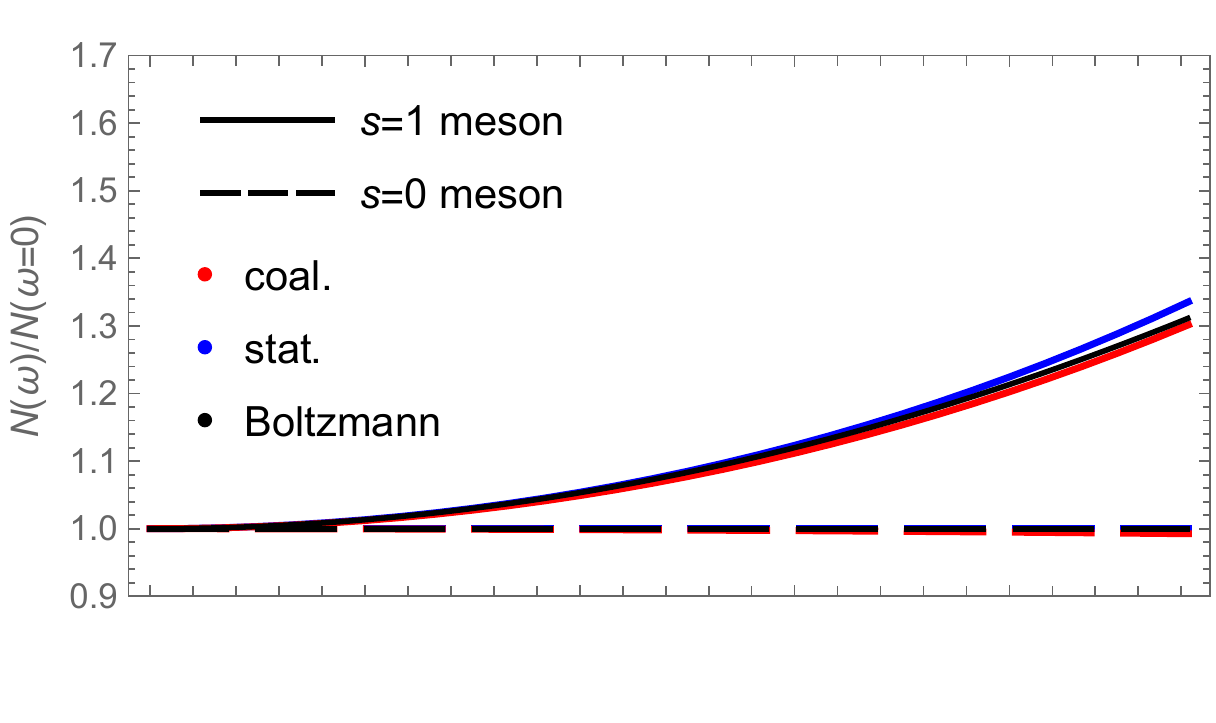} \\ \vspace*{-7mm}
\includegraphics[width=0.9\linewidth]{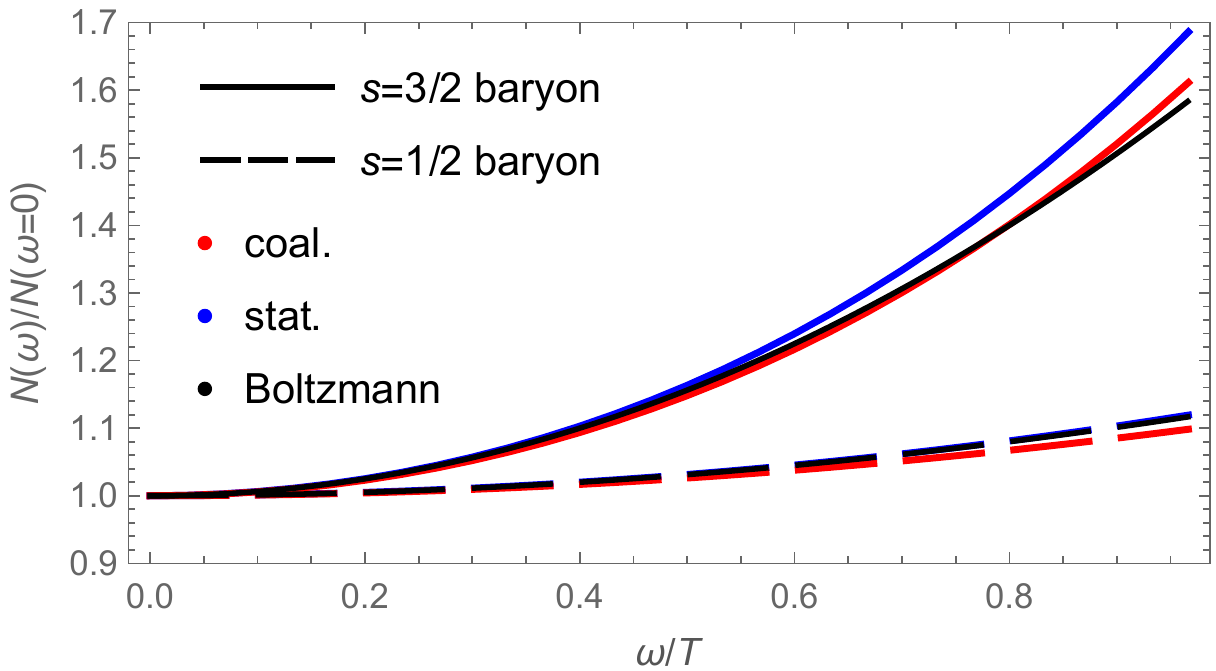}
\caption{\label{fig-1} Vorticity dependent hadron yields within the quark-coalescence (red) and statistical hadronization (blue) models, and the Boltzmann approximation (black).  Parameters are $T=160\;{\rm MeV}$, ${\bm \mu} = {\bm 0}\;{\rm MeV}$; constituent quark mass $m_{\rm q}=300\;{\rm MeV}$; and hadron mass $m_{\rm h}=1000\;{\rm MeV}$.  }
\end{center}
\end{figure}

{\it Vorticity dependent hadron yields}.  We numerically carried out the momentum integration in the thermal quark and hadron distributions, Eqs.~(\ref{eq-1}) and (\ref{eq-5}), to quantify spin-summed hadron yields within the two production models; see Fig.~\ref{fig-1}.  The results are in good agreement with the Boltzmann approximation, and we confirm that the two production models, indeed, consistently predict that spin-summed hadron yields increase quadratically with vorticity and that hadrons with larger spin are enhanced more strongly.  
The enhancement is a few percent for the typical local vorticity strength in HIC, $\omega/T \!=\! {\mathcal O}(0.1)$ \cite{wei18, jia17, hui17, xia18, kar17, cse19}.  This may be a detectable magnitude within the current experimental sensitivities, 
and drives us to further elaborate observable signals.

{\it Double ratio}.  One can extract the enhancement in Eq.~(\ref{eq-7}) by comparing different collision systems as the vorticity strength $\omega$ depends on the centrality and collision energy \cite{sta17, sta18, sta07}.  However, the signature of the vorticity needs to be distinguished from the changes of the other thermodynamic parameters $T$ and ${\bm \mu}$. To cancel such contaminations, we propose to measure the double ratio: 
\begin{align}
	D_{a,b;1,2} := \frac{N_{a}(T_1,{\bm \mu}_1,\omega_1)/N_{a}(T_2,{\bm \mu}_2,\omega_2)}{N_{b}(T_1,{\bm \mu}_1,\omega_1)/N_{b}(T_2,{\bm \mu}_2,\omega_2)},  \label{eq-8}
\end{align}
where the parameters in the two different collision systems are labeled with $i\!=\!1,2$.  The vorticity effect is signaled as the deviation from unity when we measure the two hadron spices ${\rm h} \!=\! a, b$ carrying distinct spin sizes.

\begin{figure}
\begin{center}
\includegraphics[width=0.9\linewidth]{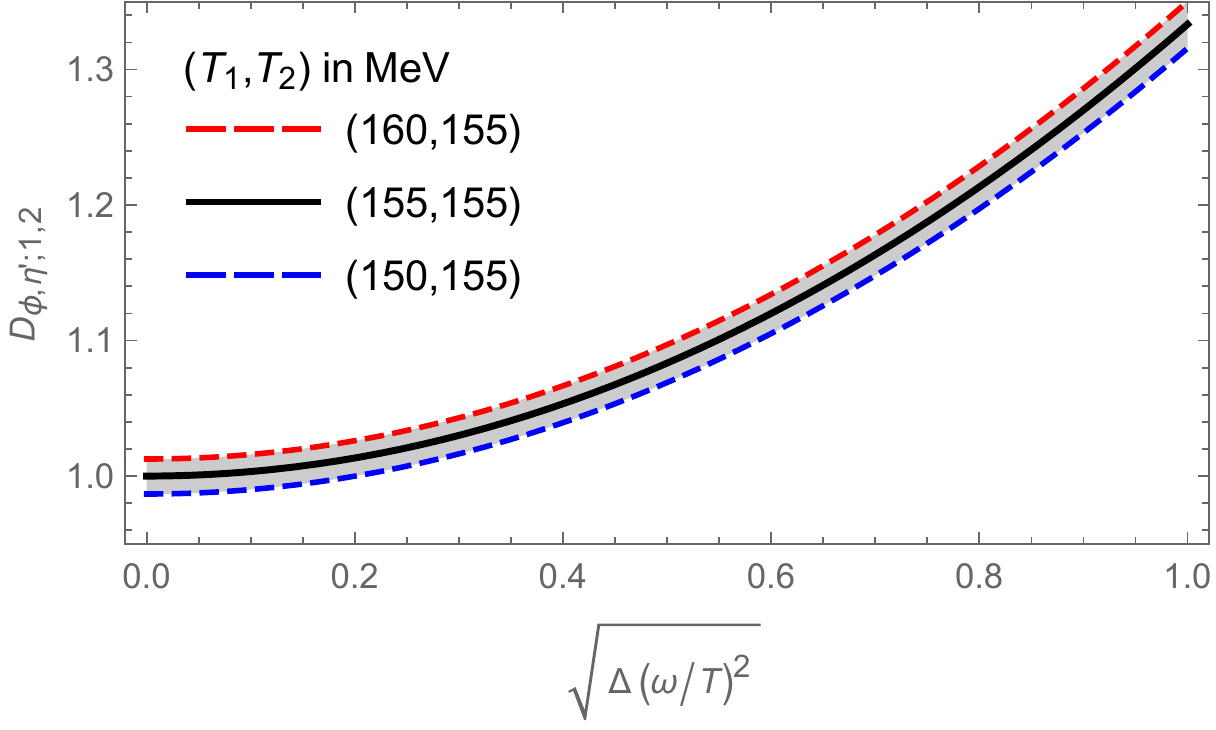}
\caption{\label{fig-2} Double ratio between $\phi$ and $\eta'$ within the Boltzmann approximation (\ref{eq-9}).  The gray region shows possible contamination by the temperature difference estimated as $T_1 = T_2 \pm 5\;{\rm MeV}$ with fixed $T_2=155\;{\rm MeV}$.  }
\end{center}
\end{figure}

One can suppress the deviation due to the temperature and chemical-potential differences by choosing the pair wisely.  Deviations caused by the chemical potentials are suppressed if the pair carries the same charges ${\bm Q}_a \!=\! {\bm Q}_b$.  There may remain deviations by the temperatures, but they are suppressed if the mass difference $\Delta m\!:=\! m_a \!-\! m_b$ is sufficiently small.  In fact, one can explicitly evaluate the double ratio $D$ within the two production models, and finds the same result in the Boltzmann limit: 
\begin{align}
	\!D_{a,b;1,2} \!\sim\! {\rm e}^{\frac{ \Delta m  \Delta T}{T_1T_2}} \!\left[ 1 \!+\! \frac{s_a(s_a\!+\!1)\!-\!s_b(s_b\!+\!1)}{6} \Delta \!\left( \frac{\omega}{T} \right)^{\!2} \right] \!, \label{eq-9}
\end{align}
where $\Delta T \!:=\! T_1 \!-\! T_2$ and $\Delta(\omega/T)^2 \!:=\! (\omega_1/T_1)^2\!-\!(\omega_2/T_2)^2$, and ${\mathcal O}((\omega_i/T_i)^4, T_i/m_{\rm h})$ terms are neglected.  To compute $D$ within the quark-coalescence model, we need to fix the overall normalization factor $N^{\rm coal}(\omega\!=\!0)$ in Eq.~(\ref{eq-2}).  We assume $N^{\rm coal}(\omega\!=\!0) \!=\! N^{\rm stat}(\omega\!=\!0)$ (by, e.g., tuning the hadron size in the quark-coalescence model \cite{exh17}), so that the quark-coalescence model is consistent with the statistical hadronization model in the absence of vorticity.  Equation~(\ref{eq-9}) indicates that $D$ is insensitive to the chemical potentials, and that the contamination by $\Delta T$ is negligible if $\Delta m$ is sufficiently small $\Delta m \Delta T / T_1 T_2 \!\lesssim\! \Delta (\omega/T )^2 \!=\! {\mathcal O}(1\;\%)$.  For example, $\Delta T$ is almost insensitive to centrality at RHIC $|\Delta T| \!\lesssim\! 5\;{\rm MeV}$ \cite{raf05}, so that hadron pairs with $\Delta m \!\lesssim\! 100\;{\rm MeV}$ can clearly show the vorticity effects.  A promising pair is $\phi(1020)$ and $\eta'(958)$, which is advantageous not only because $\Delta m$ is relatively small, but also because feed-down effects are negligible and their decay width is narrow.  In Fig.~\ref{fig-2}, we show the sensitivity of the double ratio to the vorticity strength with possible changes due to $\Delta T$.

Even if $\Delta m$ of some pair is not small, one can avoid the contamination by $\Delta T$ by additionally measuring a yield of another hadron $a'$ which has the same quantum numbers ${\bm Q}_a \!=\! {\bm Q}_{a'}$ and $s_a\!=\!s_{a'}$ as the hadron $a$ except for the mass $m_a \neq m_{a'}$.  For example, $\eta(548)$ is a possible choice for the pair $(\phi, \eta')$.  Since $N_{a}(T_i,{\bm \mu}_i,\omega_i)/N_{a'}(T_i,{\bm \mu}_i,\omega_i) \!\sim\! (m_a/m_{a'})^{3/2}\exp[ -(m_a\!-\!m_{a'})/T_i ]$ within the Boltzmann approximation, we can re-express the temperatures in the double ratio $D$ in terms of the yield of the hadron $a'$ as
\begin{align}
	\!\!\!D_{a,b;1,2}	
		\!\sim\! D_{a,a';1,2}^{\! \frac{\Delta m}{m_a-m_{a'}}} \!\!\!\left[ 1 \!+\! \frac{s_a(s_a\!+\!1)\!-\!s_b(s_b\!+\!1)}{6} \Delta \!\left( \frac{\omega}{T} \right)^{\!2} \right] \!\! .  
\end{align}
Therefore, one may directly access the local (thermal) vorticity strength $\Delta (\omega/T)^2$ without suffering from the contamination by $T$ or ${\bm \mu}$ through a measurement of $N_{\rm h}(T_i,{\bm \mu}_i,\omega_i)$ with ${\rm h}\!=\!a,a',b$ and $i\!=\!1,2$.

{\it Summary and discussions}.  We have discussed effects of vorticity on hadron yields in HIC.  We have extended the quark-coalescence and statistical hadronization models by including vorticity as a new parameter to characterize the QGP.  Based on these models, we have shown that hadron yields are enhanced by vorticity and that the enhancement is (i) the order of ${\mathcal O}((\omega/T)^2)$; (ii) independent of the direction of vorticity; (iii) larger for hadrons with larger spin; and (iv) less affected by the hadron production mechanisms for the reasonable vorticity strength estimated for HIC.  We have also proposed that the double ratio of distinct hadron species such as $(\phi, \eta')$ may be a good observable to directly access the local vorticity in HIC without suffering from the contamination by temperature and/or chemical potentials.

The vorticity effects on hadron yields survives even after taking account of the feed-down effects, where excited hadrons decay into lower mass hadrons.  The coefficient of $(\omega/T)^2$ is averaged over various hadron species by feeding-down, where the direct production yields and branching ratios can be given by those at zero vorticity to a good precision as long as the vorticity is the order of $\omega/T\!=\!\mathcal{O}(0.1)$.  We shall discuss more about the feed-down effects in a forthcoming publication.

Let us discuss implications of our results for HIC: 

(1) Yields of hadrons with different spins may be applied to estimate the local vorticity strength in HIC.  The local vorticity strength may reach $\omega/T\!\approx\! 0.3$ at the freeze-out \cite{kar17}, and may be controlled systematically by, e.g., centrality and collision energy \cite{sta17, sta18, sta07}.  Our results indicate that the modification to hadron yields and/or the double ratio is ${\mathcal O}(1\;\%)$, which may be measurable in experiments and hence could be used as a novel observable to estimate the local vorticity strength in HIC.

(2) In actual data analyses, hadron yields are fixed quantities obtained in experiments and are to be fitted by the model parameters $T, {\bm \mu}$, and $\omega^2$.  Since finite $\omega^2$ enhances the hadron yield on average, the existence of the strong local vorticity would result in a reduction of the coalescence/chemical freeze-out temperatures.

(3) Our developed models are extensions of the traditional quark-coalescence and statistical hadronization models without vorticity \cite{fri08, bra04}.  There are several conserved quantities such as conserved charges ${\bm Q}$, energy $E$, and angular momentum $J$, with which the QGP fluid is characterized.  By introducing vorticity, or $\omega$, one can cover all the possible intensive variables conjugate to these extensive conserved quantities (${\bm Q} \to {\bm \mu}, E \to T$, and $J \to \omega$).  In addition, if one obtains a better $\chi^2$ fit in our models than in the traditional models without vorticity, it would provide new strong evidence for the existence of the local vorticity in HIC.  This is complementary to the current spin polarization measurements, which measure particular decay modes of specific hadrons such as $\Lambda, K^{\ast0}$, and $\phi$ \cite{sta07, sta17, sta18, Abelev:2008ag, Acharya:2019vpe}.

(4) Since hadron production may be enhanced nonuniformly by the local vorticity in HIC, it may contribute to elliptic flow $v_2$ or even higher harmonics.  Odd harmonics such as triangular flow $v_3$ are less affected due to symmetry.  Thus, measurement of $v_2$ and/or the difference between even and odd harmonics may tell us more about the vorticity such as its space-time distribution.  

(5) Even in central collisions, vorticity can be generated at finite rapidity \cite{xia18}.  Since hadron yields are more strongly enhanced for hadrons with larger spin, a ratio between, e.g., $\phi$ with respect to $\eta', \eta$ may increase with rapidity.  Inversely, such a ratio may be used to extract the rapidity-profile of the vorticity in HIC.  

(6) In noncentral collisions, not only vorticity but also a strong magnetic field $eB \!=\! {\mathcal O}(m_{\pi}^2)$ is created \cite{bzd12, den12} (see Ref.~\cite{Hattori:2016emy} for a review).  The strong magnetic field may survive even at the freeze-out time due to the conductance of the QGP \cite{mcl14}.  If this is the case, the magnetic field may polarize hadrons/quarks, and then hadron yields should be modified just as vorticity does.  Since a magnetic field distinguishes electric charge, one can expect charge-dependent suppression/enhancement of hadron yields and flow $v_n$, from which one could extract information about the magnetic field just as vorticity.  This is an interesting possibility for isobar collisions at RHIC (e.g., ${\rm Ru}$ and ${\rm Zr}$), which provide roughly the same vorticity but 10\;\% difference in the magnetic field.  Hence, one could purely study magnetic-field effects from the difference in hadron yields of two isobar systems.

{\it Acknowledgments}.  The authors would like to thank Takafumi~Niida for fruitful discussions during the international molecule-type workshop at Yukawa Institute for Theoretical Physics (YITP) ``Hadron Interactions and Polarization from Lattice QCD, Quark Model, and Heavy Ion Collisions (YITP-T-18-07),'' where this work was initiated.  The authors also would like to thank participants of another YITP workshop of the same type ``Quantum kinetic theories in magnetic and vortical fields (YITP-T-19-06)'' for discussions.  H.~T. was supported by National Natural Science Foundation in China (NSFC) under Grant No.~11847206.  X.-G.~H. was supported by NSFC under Grants Nos.~11535012 and 11675041. S.~H.~L. was supported by Samsung Science and Technology Foundation under Project No. SSTF-BA1901-04.  S.~C. was supported by the National Research Foundation of Korea (NRF) grant funded by the Korea government (MSIP) (No.~2018R1A5A1025563), and (No.~2019R1A2C1087107).  D.-L.~Y. was supported by Keio Institute of Pure and Applied Sciences (KiPAS) project in Keio University and the Grant-in-Aid for Early-Career Scientists (JSPS KAKENHI Grant No. JP 20K14470).  P.~G. is supported by the Grant-in-Aid for Early-Carrier Scientists (JSPS KAKENHI Grant No. JP18K13542), Grant-in-Aid for Scientific Research (C) (JSPS KAKENHI Grant No. JP20K03940), and the Leading Initiative for Excellent Young Researchers (LEADER) of the Japan Society for the Promotion of Science (JSPS).  A.~M. was supported by JSPS KAKENHI Grant No. JP19K14722.  A.~O. was supported by JSPS KAKENHI Grant Nos. JP19H05151 and JP19H01898.  K.~H. is partially supported by Grant-in-Aid for Scientific Research (C) (JSPS KAKENHI Grant No. JP20K03948).

\end{document}